\documentclass[english,american]{revtex4-2}
\usepackage[T1]{fontenc}
\usepackage[utf8]{inputenc}
\usepackage{color}
\usepackage{float}
\usepackage{amstext}
\usepackage{graphicx}
\usepackage{wasysym}
\usepackage{esint}

\makeatletter
\usepackage{lipsum}
\usepackage{lineno}

\makeatother

\usepackage{babel}
\begin{document}
\title{{Exploring nonlinearities in a positive ion-negative
ion (PINI) plasma: can other processes mimic debris-induced effects?}}
\author{Hitendra Sarkar and Madhurjya P. Bora}
\email{mpbora@gauhati.ac.in}
\affiliation{Physics Department, Gauhati University, Guwahati 781014, India}
\begin{abstract}
In this work, an analysis of nonlinear waves and {structures}
induced by an external charged debris in a positive ion-negative ion
(PINI) plasma is presented.{{} The results obtained
are compared with findings from available experiments involving PINI
plasma}. The process of formation of different nonlinear structures
is examined theoretically through a forced {Korteweg-de} Vries (fKdV)
equation, which is also verified with a multi-fluid flux-corrected
transport simulation {code (\emph{m}FCT)}. {Various}
processes which are responsible for different nonlinear {waves
and} structures excited by differently charged external debris are
pointed out. This work also points out the similarities in different
nonlinear structures {excited} by an external charged
debris and the underlying processes{{} (this work) }and
those {observed experimentally through processes that
do not involve any external debris.}
\end{abstract}
\maketitle

\section{Introduction}

Nonlinear waves ubiquitously drive the dynamics of nature's most intricate
systems, shaping complex phenomena from plasma instabilities to oceanic
rogue waves in ways both beautiful and unpredictable. Plasma solitons{{}
and shocks} remain prime examples of such structures that govern the
energy transport and wave dynamics {both in} laboratory
and astrophysical environments. Solitons {and} solitary waves can be
considered as localized compressions or rarefactions of plasma densities
arising out of the interplay between nonlinearity and dispersion.
On the contrary, shock waves are usually associated with some dissipation
mechanisms such as kinematic viscosity, Landau damping, and collision
among ions and neutrals along with inherent nonlinearity, dispersion,
and weak dissipation. During the nonlinear evolution of ion-acoustic
wave (IAW) in a dissipative plasma, the leading edge of a propagating
wave gets steepened as dissipation dominates over dispersion and a
shock front is formed. In contrast to this, in a collision-less dispersive
fluid, {shock waves} {can form} without
dissipation {and} often have an oscillatory signature
and are known as dispersive shock waves (DSWs) \citep{dsw2,dsw1,dsw3,dsw4}.
Nevertheless, the nature of these nonlinear structures is largely
determined by the background plasma conditions, constituents, and
the {associated }physical parameters.

The characteristics of{{} ion-acoustic} solitons or
{shock waves} in a usual {electron-ion
($e$-$i$)} plasma are significantly different {from}
that of a multi-ion plasma {such as} a positive ion-negative
ion (PINI) plasma{{} \citep{pini3,pini2,pini4,pini1}}.
In fact, an $e$-$i$ plasma can only support compressive solitons
with positive electrostatic potential. However, with an addition of
even a small amount of one or more extra ion species to the plasma,
drastically changes the dynamics of the plasma. It is observed that
in a multicomponent plasma, the {IAW} propagation
is mainly influenced by the lighter ions and changes in their concentrations
and temperatures may lead to substantial modification of the wave
structure via the plasma potential \citep{White1972,Nakamura1985,nakamura1999}.
If the added ion species is negatively ionized, then after a critical
concentration of these negative ions, the plasma potential can even
change its polarity to form a negative or rarefactive soliton \citep{Nakamura1985,nakamura1999}.
When an ion beam is injected into such a plasma, beam-plasma interaction
makes the evolution {even} more complex in terms of
nonlinear wave structures compared to a simple plasma \citep{nakamura1999,sharma2010,bailung2010}.
For this reason, investigations on the plasmas with negative ions
as one of the constituents {have} always been a topic of interest. Plasmas
in astrophysical and space environments such as cometary tails, planetary
ionospheres, and interstellar and molecular clouds usually are inherently
multi-species with two or more than two types of ions and offer a
rich platform for studying a variety of wave motions with subtle intricacies
\citep{pfaff1998,Hirahara2023}. These multi-species plasmas with
negative ions are also easily produced in a laboratory and are well
studied owing to their vital role in major scientific and technological
applications such as plasma processing or fusion plasmas \citep{pini1}.{{}
However, the response of a PINI plasma to an external charged debris
is a topic yet to be explored fully. Moreover, as we explain in the
next subsection, we raise a fundamental question about plasma-debris
interaction -- can other nonlinear processes mimic debris-induced
nonlinearities? As we shall see, under certain circumstances, this
is indeed true.} So, we believe that an investigation of nonlinear
wave propagation in a PINI plasma embedded with an external charged
debris is definitely a topic worth studying, {which
should help us in understandings about the intricacies of the processes
leading to formation of these nonlinear structures in multicomponent
plasma.}

{Earlier and recent studies by several authors \citep{sen2015,kumar2016,truitt2020,das2024}
on phenomena involving debris-induced nonlinearities identified the
compressive structures in the precursor and wake regions of debris
as well as pinned solitons at the location of the debris. This work
shows that rarefactive structures can also be excited by debris, owing
to the presence of negative ions. Notably, for the same debris velocity,
the morphology of the excited waves in the three regions varies significantly
with the concentration of negative ions. The new findings also show
certain qualitative similarities with structures observed in other
experimental processes suggesting a broader universality in the underlying
nonlinear wave excitation mechanisms.}

\subsection{Can other {processes} mimic debris-induced nonlinearities?}

Effect of presence of external debris in a flowing plasma has definitely
gathered attention in recent years, especially as awareness and knowledge
about interaction of space debris in low-earth orbits with ionospheric
plasmas have come to the forefront of space research \citep{sen2015,jaiswal2016,kumar2016,truitt2020,chakraborty2022,sarkar2023,Kumar2024,das2024}.
Though science of detection of space debris through debris-induced
plasma interaction is still quite speculative, we are definitely learning
more and more about it, as research intensifies in this area. In this
context, we would like to raise a fundamental question about the similarities
of the dynamics and structures of nonlinear waves in a plasma in general,
to those induced by external charged debris. We do hope that we shall
be able {to }clear some issues about this question
through the present work.

In many experiments involving beam-plasma interactions, the beam is
produced \emph{in situ}, which leaves the plasma quasi-neutral. Different
nonlinear structures produced due to these interactions have interesting
dynamical signatures. In such cases, if the beam velocity $v_{b}$
is $\ll v_{s}$ ($v_{s}${is} the effective sound
velocity), the propagating IAW does not \emph{see} the Debye-scale
disturbances created by the beam. This is analogous to what is known
as `plasma approximation', which requires $(k\lambda_{D})^{2}\ll1$
{to be applicable}, where $k$ is the wave number
of the IAW and $\lambda_{D}$ is the electron Debye shielding length.
We argue that so long as $v_{b}$ remains considerably smaller than
$v_{s}$, the plasma quasi-neutrality is maintained. However as $v_{b}\sim v_{s}$,
the beam-plasma dynamics start getting affected by the IAW. And when
$v_{b}\gg v_{s}$, the situation becomes quite different and the beam
dynamics will be completely \emph{detached} from the ion-acoustic
dynamics. If we now consider the ion-acoustic time scale $\tau_{{\rm IA}}$,
it comes out to be $\sim0.01-0.5\,{\rm milli}\,{\rm second}$ for
a typical laboratory device. On the other hand, the electron-ion collision
time $\tau_{ei}$ is quite large $\sim5\,{\rm milli}\,{\rm second}$
for such parameters. As a result, when $v_{b}\gg v_{s}$, there is
no way that thermalization can occur between the beam and the background
plasma and the background plasma should \emph{see} the beam as an
external charged perturbation (debris) and the beam-plasma dynamics
should closely mimic the dynamics observed in plasma-debris interaction.
In subsequent sections, we show that this is indeed true and can be
proven quite reasonably. We also see similarities among the nonlinear
structures produced by externally induced perturbation in plasma to
that of debris-induced perturbation. In this regard debris-induced
perturbations can be thought to be quite commonplace in many experimental
arrangements.

In this work, we explore different nonlinear structures excited by
a moving charged debris in a multicomponent PINI plasma and also look
at the similarities of their characteristics with {different}
experimental results. The complete plasma system is modeled using
a forced Korteweg--de Vries (fKdV) equation and results from theoretical
analysis are verified using a 1-D multi-fluid flux-corrected transport
simulation {code (\emph{m}FCT)} \citep{sarkar2023}. The
paper is organized as follows. The detailed plasma model and the governing
equations are described in Section II. Section III contains the description
of {the} fKdV dynamics and interpretation of its results.
{The} results {of Section III are
verified using} FCT simulation{{} in Section IV}. In
Section V, we briefly address the variation of nonlinearity with debris
velocity and in Section VI, we conclude.

\section{Plasma model and governing equations}

The model we are going to consider is a 1-D warm positive ion-negative
ion (PINI) {dissipation-less} plasma with an external
charge debris embedded into it. The relevant equations are the continuity
and momentum equations for positive and negative ions. The equations
are closed by the Poisson equation. The electrons are considered to
be Boltzmannian. The equations are given by
\begin{eqnarray}
\frac{\partial n_{\pm}}{\partial t}+\frac{\partial}{\partial x}(n_{\pm}v_{\pm}) & = & 0,\label{eq:cont}\\
n_{+}\frac{dv_{+}}{dt} & = & -\sigma\frac{\partial n_{+}}{\partial x}-n_{+}\frac{\partial\phi}{\partial x},\label{eq:n+}\\
n_{-}\frac{dv_{-}}{dt} & = & -\frac{\sigma}{\mu}\frac{\partial n_{-}}{\partial x}+\frac{n_{-}}{\mu}\frac{\partial\phi}{\partial x},\label{eq:n-}\\
\frac{\partial^{2}\phi}{\partial x^{2}} & = & n_{e}-\delta n_{+}+(\delta-1)n_{-}-\rho_{{\rm ext}},\label{eq:pois}
\end{eqnarray}
where $(n_{\pm},v_{\pm})$ are positive and negative ion densities
and velocities {respectively and} $\phi$ is the plasma
potential. The quantity $\sigma=T_{i}/T_{e}$ is the ratio of ion
temperature to that of the electrons. Here, we have taken the temperatures
of positive and negative ions to be equal $T_{i}=T_{+}=T_{-}$. The
mass ratio of negative to positive ions is denoted by $\mu=m_{-}/m_{+}$
and $\delta=n_{+0}/n_{e0}$ is the ratio of the equilibrium positive
ion density to the equilibrium electron density. In order to be able
to compare our analytical results with existing laboratory experiments,
we have assumed $m_{+}>m_{-}$, which is mostly the case in case of
experiments \citep{Nakamura1985,nakamura1999,sharma2010,bailung2010,pathak2025}.
In the above equations, the densities are normalized by their respective
equilibrium values and potential is normalized by $(T_{e}/e)$ {with
temperature expressed in energy unit.} Time is normalized by $\omega_{i}^{-1}$,
with $\omega_{i}=\sqrt{n_{e0}e^{2}/(m_{+}\epsilon_{0})}\equiv\delta^{-1/2}\omega_{p+}\equiv(\delta-1)^{-1/2}\omega_{p-}$,
where $\omega_{p\pm}$ are the respective positive ion and negative
ion plasma frequencies. While length is normalized by electron Debye
length, velocities are normalized by the ion sound speed $c_{s}=\sqrt{T_{e}/m_{+}}$,
determined by the mass of the heaviest species. The quantity $\rho_{{\rm ext}}\equiv\rho_{{\rm ext}}(x-v_{d}t)$
is the normalized charge density of the external debris with $v_{d}$
as the normalized velocity of the external debris. The overall charge
neutrality of the PINI plasma without external debris is ensured through
the relation
\begin{equation}
n_{e0}+n_{-0}=n_{+0}.
\end{equation}
The normalized electron density is given by $n_{e}=e^{\phi}$. {It should however be noted that with an external charged debris, the above quasi-neutrality condition \emph{does not} hold good { at the site of debris} and this is also the precise reason why the system becomes non-integrable (see the next Section for more discussion). {We also note that given the case for ion-acoustic solitons or solitary waves, instantaneously there may be a violation of the Boltzmannian electrons when $e\phi\sim m_e v_{{\rm th}e}$, where $v_{{\rm th}e}$ is the electron thermal velocity. However, considering the overall IA timescale $\tau_{\rm IA}\gg\omega_{pe}^{-1}$, the electron plasma frequency, we can justify the electron equilibrium to be quasi-static even though the ion response is nonlinear. This can even be justified from kinetic theory as well by using the stationary Vlasov equation for electrons (dimensional)
\begin{equation}
v_e\frac{\partial f_e}{\partial x}-\frac{e}{m_e}\frac{\partial\phi}{\partial x}\frac{\partial f_e}{\partial v_e}=0,
\end{equation}
assuming  $\partial f_e/\partial x\sim0$ in the IA timescale, which leads to the Boltzmannian distribution on integration \citep{boltz}
\begin{equation}
n_e=\int f_e\,dv_e\propto\exp\left(\frac{e\phi}{T_e}\right).
\end{equation}
In the above expressions, $v_e,m_e$ are respectively electron velocity and mass. $f_e$ is electron velocity distribution function, which is assumed to be Maxwellian.
}}

{An explanation about the `dissipation-less' assumption
is given in Section III (just before Section III-A), where it is more
appropriate to discuss it with reference to the KdV equation.}

\subsection{The morphology of debris-induced nonlinearities}

Before we investigate the nonlinear waves and structures induced by
a charged debris in a PINI plasma, it is of relevance to discuss a
bit about the fundamental physics of these nonlinearities in an $e$-$i$
plasma. We now know that the response of an $e$-$i$ plasma to an
embedded charged debris very well depends on the nature of the charge
of the debris \citep{sarkar2023}. It has been \emph{only} recently
shown that the formation of bright pinned solitons in the ion-acoustic
regime due to the presence of negatively charged debris is basically
a manifestation of the trapped ions in phase-space vortices formed
due to ion-ion counter streaming instability (IICSI) \citep{das2024}.
Besides, proper spatial resolution of the pinned solitons requires
a certain relative velocity between the debris and the plasma, though
higher relative velocities may cause the pinned solitons to disappear
\citep{kumar2016,das2024}. In contrast, a positively charged debris
creates an ion hole, exciting an ion-acoustic wave (IAW), which propagates
away from the site of the debris. However, when there is a relative
motion between the debris and the plasma, the IAW becomes a dispersive
shock wave (DSW) in the precursor region \citep{sarkar2023}.
{We should, however be careful, while referring to these so-called `pinned' solitons. 
Though they are referred to as `solitons', they are characteristically different
from true solitons as a true soliton does not evolve in time, retains its shape,
and is asymptotically stable,
which is produced by an integrable system. On the other hand, pinned solitons, produced
by external charged debris are
\emph{not} true solitons as they are not asymptotically stable and produced
by non-integrable systems.}

Following the above explanation, we expect that the response of a
PINI plasma to external charged debris should also be characteristically
different for positively and negatively charged debris. We further
expect that for $\delta\sim1$, the response should be similar to
that in an $e$-$i$ plasma.

\subsection{{Mathematical reduction}}

{We note that the above plasma model can be analyzed
both mathematically and numerically. Typically, in {this} class of
problems, one uses a scale transformation and reduces it to a pseudo
potential (or Sagdeev potential) form \citep{Sagdeev} through the
Poisson equation, which guarantees a soliton solution, though not
necessarily a `${\rm sech}$'-type solution as one would have obtained
through a KdV equation \citep{kdv}. The other two approaches would
be a reductive perturbation analysis to reduce the system to a KdV
equation or a nonlinear Schr\"odinger equation (NLSE) \citep{nlse}.
However, with an external charged debris term, the model is simply
not amenable to pseudo-potential analysis and that leaves us with
other two options.}

{Employing the reductive perturbation method with
the external charged debris term, we end up with either a forced-KdV
(fKdV) or a forced-NLSE equation. At this point, we should emphasize
that numerically we are going to a use flux-corrected transport (FCT)
method (as mentioned before) to simulate the dynamics, which is very
efficient in detecting sharp discontinuities that can result from
the debris-induced perturbation. Also the FCT formalism does not employ
any approximations and solves the full set of equations, which in
principle should provide us with both {KdV-like} as well as {NLSE-like}
modulated wave solutions. However, the single hump perturbation that
we expect from an external charged debris, the numerical solution
is most likely to favor {KdV-like} solution which is also closer to
the real-world and observable dynamics. In what follows, we shall
therefore analyze the above model mathematically, using the fKdV equation
only.}

\section{Forced-KdV dynamics}

We shall now see, what forced-KdV (fKdV) dynamics has to offer. The
fKdV equation is derived using the usual reductive perturbation theory
by using the following expansions, assuming a static and neutral equilibrium.
The external charged debris is introduced as a second order perturbation.
The expansions are 
\begin{eqnarray}
n_{\pm} & \simeq & 1+\varepsilon n_{\pm}^{(1)}+\varepsilon^{2}n_{\pm}^{(2)}+\cdots,\label{eq:nexp}\\
v_{\pm} & \simeq & \varepsilon v_{\pm}^{(1)}+\varepsilon^{2}v_{\pm}^{(2)}+\cdots,\\
\phi & \simeq & \varepsilon\phi^{(1)}+\varepsilon^{2}\phi^{(2)}+\cdots,\label{eq:phiexp}\\
\rho_{{\rm ext}} & = & \varepsilon^{2}\rho_{{\rm ext}}^{(2)},
\end{eqnarray}
where $\varepsilon$ is the expansion parameter, which {should be} $\ll1$.
{The expansion parameter $\varepsilon$ is \emph{not} a physical parameter of the system in the sense that it does not come from measurement of any physical quantity and it needs to be interpreted in the context of the numerical solution of the resultant fKdV equation (see Appendix A for a possible definition).}
{One question related to the order of perturbation
for the external charged debris may naturally arise here. In this
context, two points to be noted here -- that we {usually} use the KdV-{type} equation
for weak nonlinearity  and how strong a perturbation an external
charged debris may induce in the bulk plasma. Our assumption of using
a second order perturbation for the debris is quite in line with the
weak nonlinearity assumption on the basis of which the {fKdV} equation
is derived. As far as the strength of perturbation is concerned, we
shall see in subsequent sections that the experimentally observed
nonlinearities are quite close to our numerical solutions (obtained
via FCT simulation) as well as the fKdV solutions, which demonstrates
that our assumption of introducing the external charged debris as
second order perturbation is indeed consistent {(an explanation justifying this is given in Appendix A).
We should {further} note that the reductive perturbation theory essentially requires that all external perturbations must \emph{not} exist at the first order or else the theory would fail (see Appendix B for a simple explanation).}
 Besides, using a second
order perturbation for debris leaves us with the simplest form of
{the} fKdV equation.}

We now use the stretched variables $\xi=\varepsilon^{1/2}(x-Vt)$
and $\tau=\varepsilon^{3/2}t$ for space and time coordinates, where
$V$ is the dimensionless phase velocity of the IAW. Following the
usual procedure, by collecting the first order terms, we have the
following expressions for the first order variables in terms of $\phi^{(1)}$,
\begin{eqnarray}
n_{+}^{(1)} & = & \frac{\phi^{(1)}}{V^{2}-\sigma},\\
n_{-}^{(1)} & = & -\frac{\phi^{(1)}}{V^{2}\mu-\sigma},\\
v_{\pm}^{(1)} & = & Vn_{\pm}^{(1)},
\end{eqnarray}
where the phase velocity expression (or the compatibility condition)
is given by
\begin{equation}
V=\left[\frac{\alpha+\sqrt{\alpha^{2}-4\mu\sigma(2\delta+\sigma-1)}}{2\mu}\right]^{1/2},
\end{equation}
with $\alpha=(\mu+1)(\delta+\sigma)-1$. Collecting the next higher
order terms and eliminating the second order quantities, using the
Poisson equation, we finally get the fKdV equation as follows
\begin{equation}
\partial_{\tau}\phi^{(1)}+A\phi^{(1)}\partial_{\xi}\phi^{(1)}+B\partial_{\xi}^{3}\phi^{(1)}=-B\partial_{\xi}\rho_{{\rm ext}}^{(2)},\label{eq:fkdv}
\end{equation}
where nonlinear and dispersion coefficients $A$ and $B$ are given
by
\begin{eqnarray}
A & = & \left[\frac{(b+2V^{2})\delta}{b^{3}}-\frac{2(\delta-1)V^{2}\mu}{a^{3}}-\frac{(\delta-1)}{a^{2}}-1\right]B,\\
B & = & \left[2V\left\{ \frac{(\delta-1)\mu}{a^{2}}+\frac{\delta}{b^{2}}\right\} \right]^{-1},
\end{eqnarray}
and $a=(V^{2}\mu-\sigma)$ and $b=(V^{2}-\sigma).$

We note that unlike the KdV equation, the fKdV equation, in general,
is non-integrable and has to be solved numerically. So, we present
the numerical solutions of Eq.(\ref{eq:fkdv}) for a Gaussian charge
density profile for the external debris with periodic boundary conditions.{{}
Here, we consider two possible scenarios -- positively charged debris
(Section III-A) and negatively charged debris (Section III-B). For
positively charged debris, we also compare the debris-induced nonlinearity
with results from two different experiments, both of which employ
different mechanisms to excite nonlinear waves.} All the figures in
the subsequent sections are drawn in the rest frame of the debris.
We present our results with respect to two independent parameters
-- $\delta$ and $v_{d}$ for a fixed $\sigma$.

{As mentioned before, we now provide an explanation
about our `dissipation-less' assumption of the plasma model. The dissipation
in this case may come in two flavors -- through collisions and Landau
damping. There are examples of work involving multicomponent plasma
(not necessarily PINI plasma) with collisions as well as with Landau
damping term \citep{arnab}. While the inclusion of different collisional
terms in the fluid model is quite {straightforward} \citep{misra,chak},
the same is not true in the case of Landau damping. However, as shown
by VanDam \citep{vandam}, it is indeed possible to include a dissipation-like
term arising out of wave-particle resonance (Landau damping) in a
KdV-like formalism. In Section III-A.1, while comparing our fKdV solutions
with ion-beam-induced nonlinearities, we shall see that the $e$-$i$
collision time $\tau_{ei}\sim3.5\,{\rm milli}\,{\rm second}$. However,
the ion-acoustic time scale, which is the characteristic time scale
of this work, is $\tau_{{\rm IA}}\sim0.03\,{\rm milli}\,{\rm second}$.
So, we can probably safely neglect collisional dissipation. Coming
back to now Landau damping, VanDam \citep{vandam} has shown that
a dissipation-like term can be incorporated into the KdV equation
with the linear Landau damping rate $\gamma=\tilde{\gamma}|k|$ (see
Appendix C) where
\begin{equation}
\tilde{\gamma}=\frac{\lambda}{\alpha\varepsilon\sqrt{2\pi}}\left[\left(\frac{m_{e}}{m_{+}}\right)^{1/2}+\sigma^{-3/2}\exp\left(-\frac{\lambda^{2}}{2\sigma}\right)\right],
\end{equation}
with
\begin{eqnarray}
\lambda & = & 1+\frac{3}{2}\sigma,\\
\alpha & = & 2\lambda^{-2}-3\lambda^{-4}\sigma.
\end{eqnarray}
In the above expression, $k$ is the wavenumber of the associated
IAW. The expansion parameter can be taken as $\varepsilon\sim\sigma^{-3/2}e^{-1/(2\sigma)}$
\citep{vandam}. This expression is written with the assumption that
the momentum of the resultant IAW in a PINI plasma is carried primarily
by the heavier positive ions. With typical parameters used in this
work (see Section III-A.1), we find that the damping timescale $\tau_{\gamma}$
corresponding to $\gamma$ is $\sim1\,{\rm milli}\,{\rm second}$,
which is more than two orders of magnitude larger than $\tau_{{\rm IA}}$.
So, owing to the fact that $\tau_{\gamma}\gg\tau_{{\rm IA}}$, we
can safely neglect the effect of Landau damping in this work.}

\subsection{{Positively} charged external debris}

\begin{figure}[H]
\begin{centering}
\includegraphics[width=0.5\textwidth]{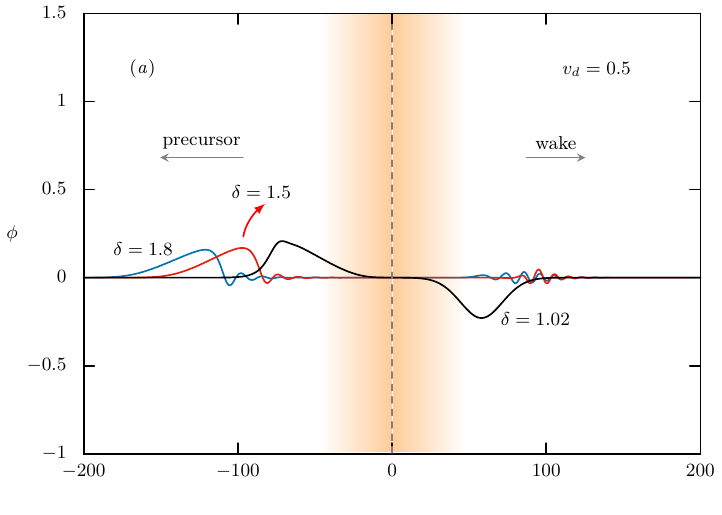}\hfill{}\includegraphics[width=0.5\textwidth]{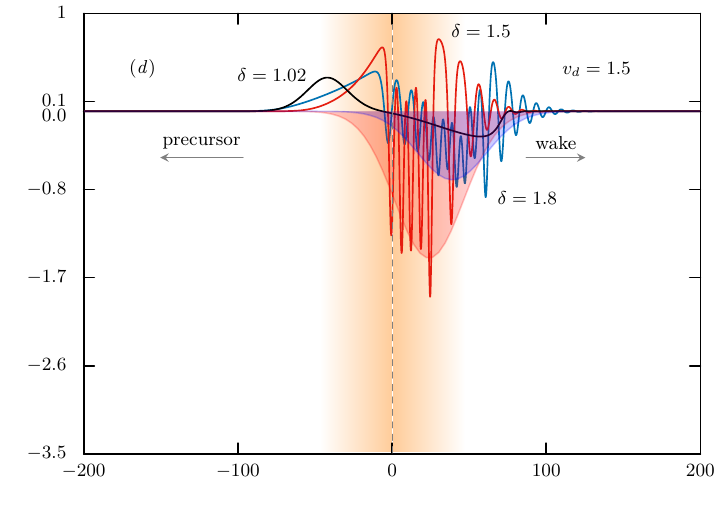}\\
\includegraphics[width=0.5\textwidth]{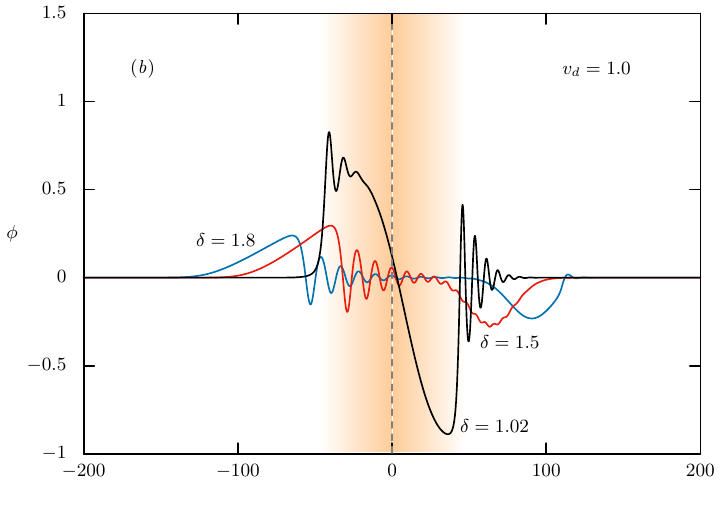}\hfill{}\includegraphics[width=0.5\textwidth]{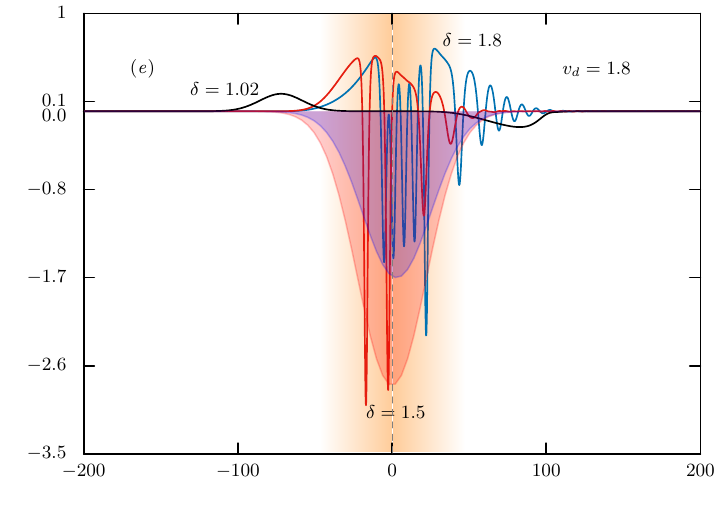}\\
\includegraphics[width=0.5\textwidth]{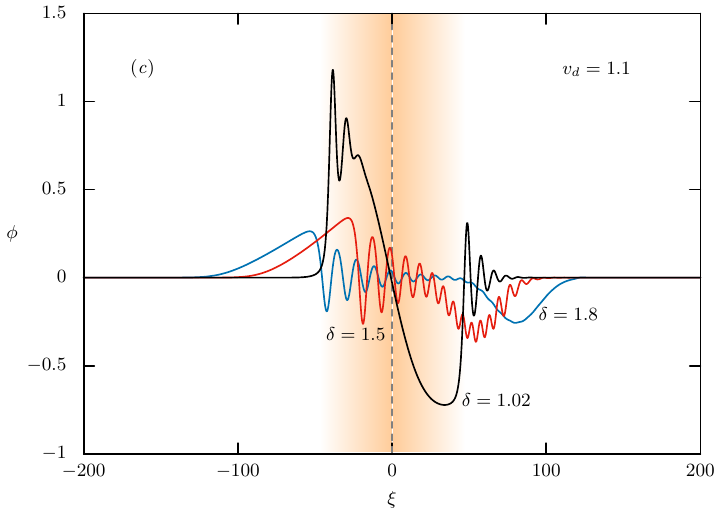}\hfill{}\includegraphics[width=0.5\textwidth]{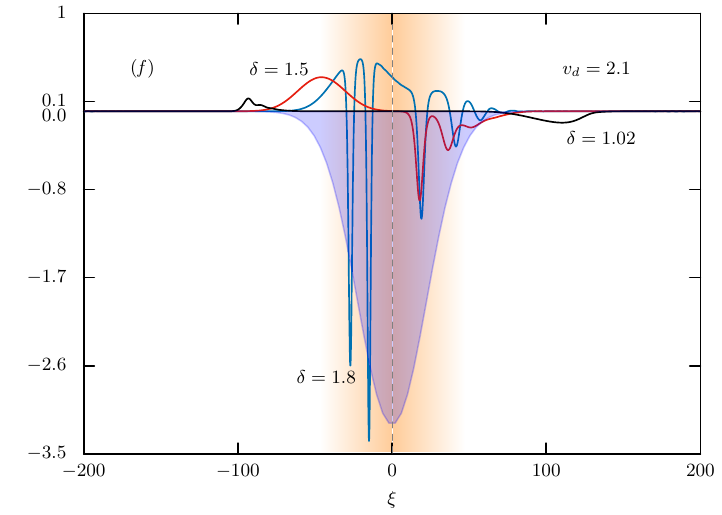}
\par\end{centering}
\caption{Nonlinear wave for a flowing PINI plasma with a positively charged
external debris. Panels (\emph{a\/})-(\emph{f\/}) correspond to debris velocities $v_d=0.5,1.0,1.1,1.5,1.8$, and $2.1$, respectively. All panels are drawn in the rest frame of the debris,
which is in the middle of a panel, denoted by a vertical dashed line.
The shaded vertical regions indicate the extent of the charged debris.
The shaded envelopes on the right hand panels indicated the envelopes
of the corresponding dark pinned solitons.}\label{fig:first}
\end{figure}

The first set of results shown in Fig.\ref{fig:first} are for debris
with a positive charge $\rho_{{\rm ext}}>0$ at the end of $\text{\ensuremath{\tau}}=100$
with $\sigma=0.1$. The figure shows the nonlinear waves in $\phi\equiv\phi^{(1)}(\xi,\tau)$.
The charge density distribution of the debris is a Gaussian-shaped distribution{{} (see below) }$\rho_{{\rm ext}}(\xi)=\rho_{0}e^{-\xi^{2}/\Delta}$,
where $\rho_{0}$ is the peak of the distribution and width is
\noindent determined
by $\Delta$ \citep{sen2015}. 
We note that the external debris is
introduced as a perturbation with $\rho_{0}$ denoting the strength.
In all the cases (unless indicated otherwise), we have used a perturbation
with {$\rho_{0}\sim0.5$} which is equivalent to a perturbation $\equiv50\%$
of the equilibrium density.
{Regarding the perturbation amplitude, we would like to note that the magnitude of perturbation of $\sim50\%$ of the equilibrium density may not look like a weak perturbation. However, when we translate it to the nonlinearity regime, the perturbation indeed has introduced a nonlinearity, which the }
\begin{figure}[H]
\begin{centering}
\includegraphics[width=0.5\textwidth]{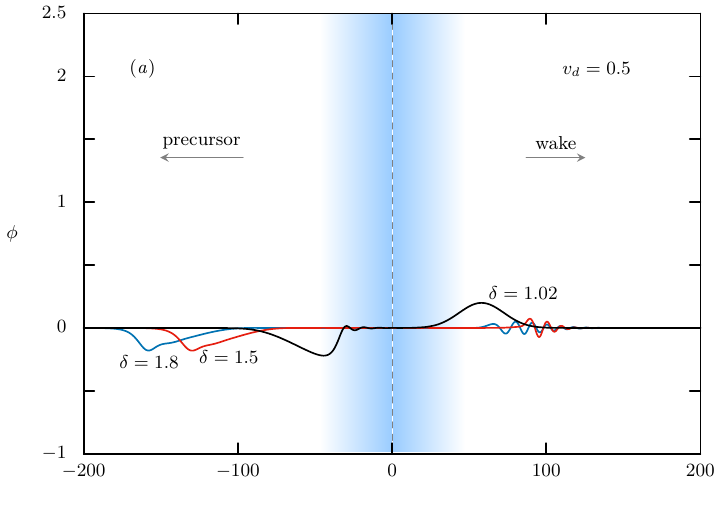}\hfill{}\includegraphics[width=0.5\textwidth]{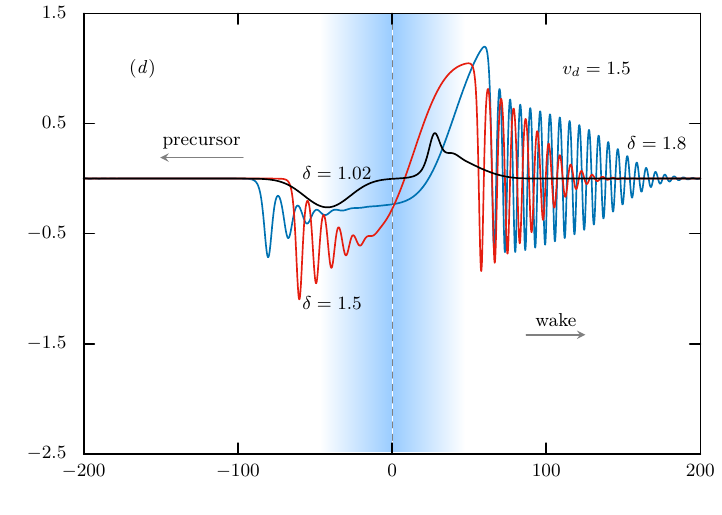}\\
\includegraphics[width=0.5\textwidth]{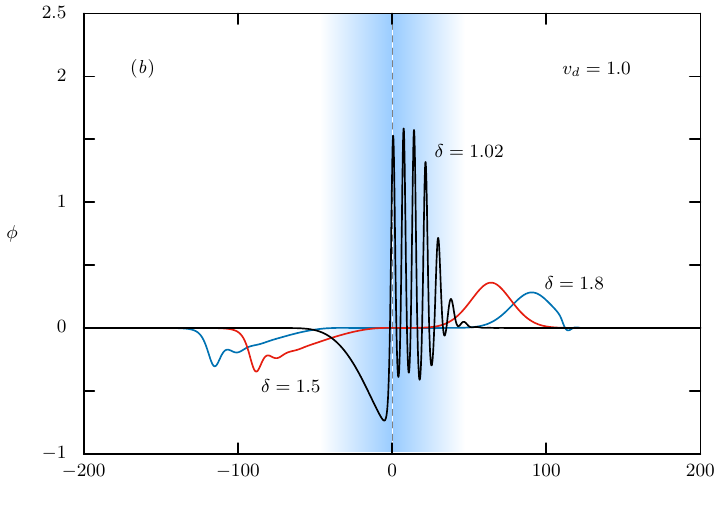}\hfill{}\includegraphics[width=0.5\textwidth]{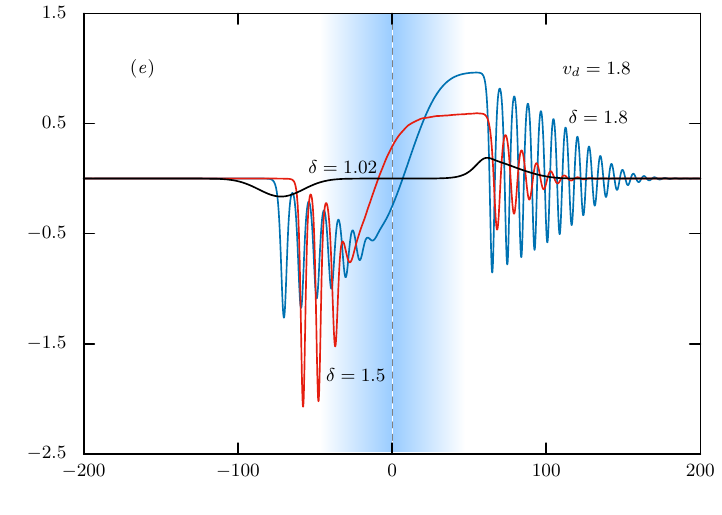}\\
\includegraphics[width=0.5\textwidth]{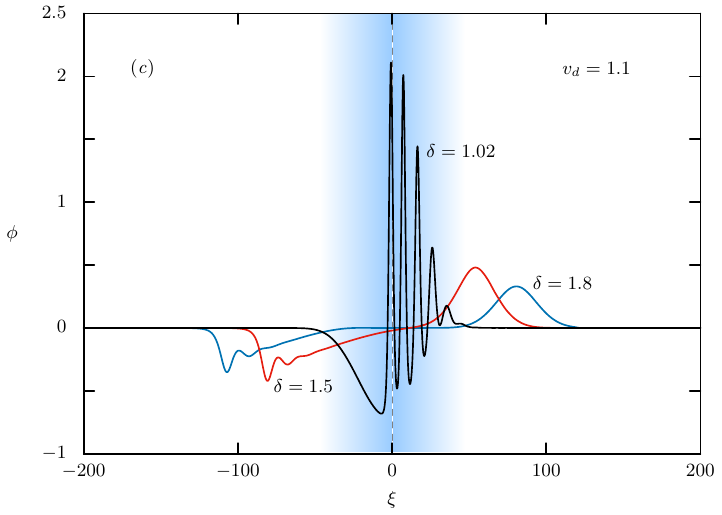}\hfill{}\includegraphics[width=0.5\textwidth]{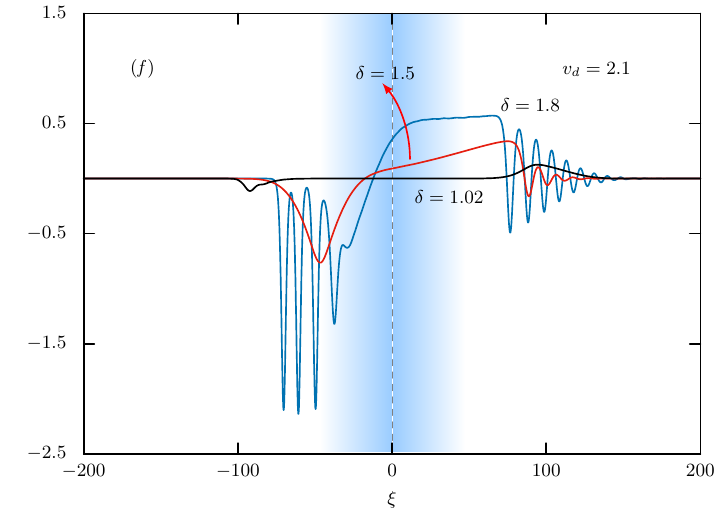}
\par\end{centering}
\caption{Nonlinear wave for a flowing plasma with a negatively charged external
debris. Panels (\emph{a\/})-(\emph{f\/}) correspond to debris velocities $v_d=0.5,1.0,1.1,1.5,1.8$, and $2.1$ respectively. All other particulars are same as in Fig.\ref{fig:first}.
The shaded vertical regions indicate the extent of the negatively
charged debris.}\label{fig:second}
\end{figure}
{\noindent fKdV equation can handle (see Appendix A).
This fact is verified through FCT simulation (see Section IV), where we have estimated the expansion parameter $\varepsilon$ from simulation to be about $\sim0.7$ with simulation results closely agreeing our fKdV results.}
In the figure, the shaded region shows
the extent of the debris charge distribution centered in the middle
of the frame indicated by the dashed vertical line. The debris move
with a velocity $v_{d}$ from right to left. 

In Fig.\ref{fig:first}(\emph{a}),
we observe that for low (subsonic) $v_{d}$, the debris excites bright
ion-acoustic {solitary wave} which move away from the debris in the precursor
region. We also observe that the velocities of these {solitary waves} increase
from being subsonic to becoming supersonic as $\delta$ increases
from $1.02$ to $1.8$. {The changing shape of the
{soliton-like structure} for increasing $\delta$ can be attributed to the increasing
negative ion density (with higher $\delta$) which tend to accelerate
from left to right as a response to the positive perturbation at the
debris site.} We {further} note that in the production
of a PINI plasma, experimentalists usually use an electronegative
gas in an $e$-$i$ plasma to create negative ions at the expense
of less warm electrons. So, an increase of $\delta$ causes an increase
in negative ion density in a PINI plasma and the increase of velocity
of the soliton can be attributed to the increase in negative ion density.
However, as debris velocity increases and becomes sonic, higher $\delta$
perturbations slow down in comparison to lower $\delta$ perturbations
which ultimately forms dark pinned solitons as debris velocity increases
more, as shown in the three panels on the right in the figure, Fig.\ref{fig:first}(\emph{d}),
(\emph{e}), and (\emph{f}).
As $v_{d}$ increases, pinned solitons
tend to form at the debris position for higher $\delta$ whereas low
$\delta$ perturbations never get to form pinned solitons and remain
predominantly a positive nonlinearity in the form of a DSW in the
precursor region.

The assumption of Gaussian-shaped
distribution for the charge density of a debris of size $\gg\lambda_{D}$
can be fairly ascertained, if the parameter $\beta_{e}\ll1$ (for
negatively charged debris), with
\begin{equation}
\beta_{e}=\frac{e^{2}}{4\pi\epsilon_{0}RT_{e}},
\end{equation}
where $e$ is the electronic charge and $R$ is the radius of a spherical
debris \citep{charging}. This result is rigorously proved for impurity
particles in a plasma using the orbit motion limited (OML) theory
\citep{oml} and numerical simulation, which can be readily interpreted
for charged debris as well. This can also be extended to debris with
positive charge. 
If we now consider the case of charged debris in
space plasma in {Low Earth Orbit} (LEO), the size of the debris can
vary from being very small to very large, but anything larger than
a few Debye length is capable of inducing nonlinear structures. Considering
typical parameters for LEO plasma \citep{leo1,leo2}, the value for
$\beta_{e}$ comes out to be $\sim10^{-6}$ even for $R\sim\lambda_{D}$.
For ions (positively charged debris), this comes out to be $\beta_{i}\sim10^{-3}$.
In laboratory experiments involving debris $\beta_{e}\sim10^{-5}$
assuming debris of size $\sim\lambda_{D}$ \citep{jaiswal2016}.

These results are entirely in conformity with what we know about formation
of debris-induced pinned solitons and DSWs. We now know that in an
$e$-$i$ plasma the pinned solitons formed due to moving negatively
charged debris is basically a manifestation of the ions trapped in
the phase space vortices formed by the IICSI \citep{das2024}, whereas
a positively charged debris induces a DSW in the precursor region.
So, for $\delta\sim1$ ({very} less concentration of negative ions),
the plasma behaves like the usual $e$-$i$ plasma and no pinned solitons
can form for a moving positively charged debris.
 However for $\delta>1$, the negative ions are attracted
by the positively charged debris causing a IICSI-like phenomena for
negative ions, leading to the formation of phase space vortices which
in turn cause formation of dark pinned solitons at the debris position.

\subsubsection{Comparison with ion-beam-induced nonlinearities}

{We} now discuss the {first} case {where
under certain circumstances,} a positive ion-beam in a PINI plasma{{}
mimics the responses of a debris-induced nonlinearities. In this experiment,}
the positive ion-beam is produced in situ by accelerating a part of
the ion population in the plasma so that overall quasi-neutrality
is maintained throughout \citep{nakamura1999}. This case is also
interesting as there are several investigations being carried out
by different authors to analyze the formation of nonlinear structures
in such plasmas \citep{sharma2010,bailung2010}.
Let us consider such a three-component PINI plasma, consisting of
positive ion $(n_{+})$, negative ion $(n_{-})$, and electron $(n_{e})$,
in the presence of a positive ion beam $(b^{+})$, as considered in
the experimental work by Nakamura \citep{nakamura1999} (N99 hereafter).
Note that the whole system is to be considered quasi-neutral,
\begin{equation}
n_{+}+n_{b+}=n_{-}+n_{e}=n_{T},
\end{equation}
which essentially requires
that the positive ion beam to be created in situ. As a result, the
formation of the positive ion beam causes an overall depletion of
the bulk positive ions. The comparative concentration of negative
ion is denoted by the quantity $r=n_{-}/n_{T}$, which can be written
in terms of $\delta$
\begin{equation}
\delta=\frac{1}{1-r}.
\end{equation}
Let us now consider the slowest possible ion-acoustic time scale which
is given by
\begin{equation}
\tau_{{\rm IA}}=f^{-1}=\left[\frac{k}{2\pi}\sqrt{\frac{T_{e}}{m_{+}}}\right]^{-1},
\end{equation}
where $k=2\pi/\lambda$ with $\lambda$ being the wavelength of the
ion-acoustic wave. In the above expression, $T_{e}$ is the electron
temperature which is $\sim1\,{\rm eV}$ and $m_{+}$ is the mass of
the heaviest ion, which is ${\rm Ar}^{+}$, so that $m_{+}\sim40m_{p}$,
where $m_{p}$ is the mass of a proton. Assuming the experimental
double-plasma device to be of length $\sim90\,{\rm cm}$ \citep{nakamura1999},
we can assume the largest possible ion-acoustic wavelength to be $\sim0.45\,{\rm m}$.
With these parameters, we have $\tau_{{\rm IA}}\sim0.03$ milli second.
If we now consider the electron-ion collision time $\tau_{ei}$, we
have
\begin{equation}
\tau_{ei}=\frac{6\sqrt{2}\pi^{3/2}\epsilon_{0}^{2}m_{e}^{1/2}T_{e}^{3/2}}{n_{e}e^{4}\,\ln\Lambda},
\end{equation}
where $\ln\Lambda$ is the Coulomb logarithm $\sim10-12$. For $n_{e}\sim10^{7}\,{\rm cm}^{-3}$
\citep{nakamura1999}, we have $\tau_{ei}\sim3.5$ milli second. We
now argue that as $\tau_{ei}\gg\tau_{{\rm IA}}$, it can be safely
assumed that in the ion-acoustic timescale, the plasma will not be
able to thermalize. As a result, a positive ion beam propagating through
the plasma will not be able to thermalize with the bulk plasma and
for all practical purposes, the beam will behave as an external debris.
It has been observed in N99 that below a certain critical value for
beam velocity, increasing $r$ (higher $\delta$) causes the nonlinearity
to become negative from positive. \foreignlanguage{american}{However,
beyond the critical value of the beam velocity, the nonlinearity remains
positive irrespective of the initial pulse (which is reported only
at higher $\delta$ in N99).} In our case \foreignlanguage{american}{{with
debris }}as well, we see a similar behavior as evident from \foreignlanguage{american}{the
panels in the left column of the figure i.e.\ Fig.\ref{fig:first}(\emph{a}),
(\emph{b}), and (\emph{c}). We can see that when debris velocity is
$\apprle1.1$, for low $\delta$ $(\sim1)$, there is a steepening
in the leading edge indicating positive nonlinearity while for higher
$\delta$, the steepening is on the falling edge, which indicates
negative nonlinearity. We note that the `positive nonlinearity' in
the case of a compressive or bright soliton-like structure is referred with respect
to the nonlinear term $A\phi^{(1)}\partial_{\xi}\phi^{(1)}$ {[}see
Eq.(\ref{eq:fkdv}){]} in a usual KdV equation (without the debris
term), which is responsible for steepening of the leading edge of
a soliton-like structure. On the other hand, when the steepening occurs in the falling
edge, it is usually termed as `negative nonlinearity'. When debris
velocity increases {[}Fig.\ref{fig:first}(\emph{f}){]}, the negative
nonlinearity which was observed for say $\delta=1.5$ and with low
$v_{d}$, decreases and {gets} balanced by dispersion giving rise to
a bright KdV soliton-like pulse. These results are also confirmed through our
FCT simulation, which are being presented in {Section
IV}.}

We should however be cautious not to compare the entire experimental
results of N99 with the results obtained in this work, as the external
charged debris in this work is a localized charged perturbation with
no assumption of quasi-neutrality, while in N99 and other similar
experimental works, the beam is an extended structure with definitive
wake region. So, we can probably compare only the leading edge dynamics
of a beam-plasma interaction with the precursor region of debris-induced
nonlinearity.

\begin{figure}
\begin{centering}
\includegraphics[width=0.5\textwidth]{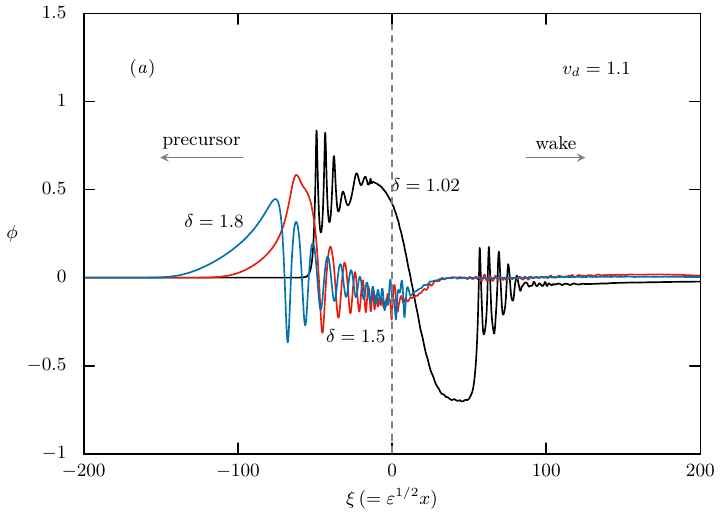}\hfill{}\includegraphics[width=0.5\textwidth]{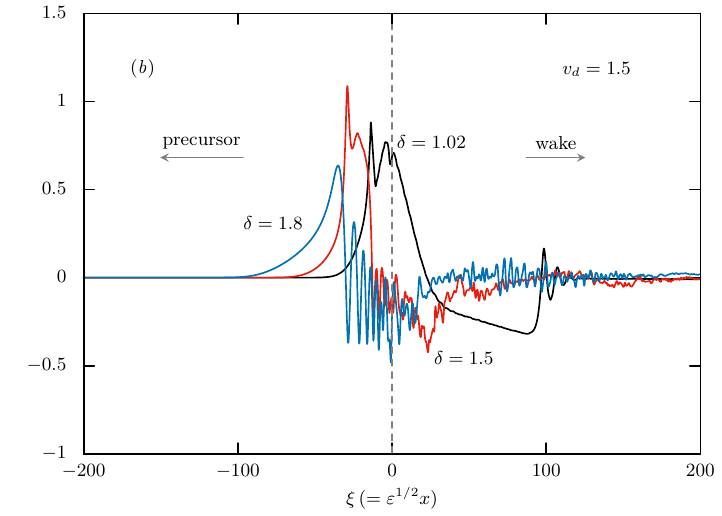}
\par\end{centering}
\caption{FCT simulation results of propagation of nonlinear wave in a PINI
plasma with {an} embedded external positively charged debris. These
results are equivalent to the theoretical cases obtained through fKdV
solution shown in Fig.\ref{fig:first}. The left panel (\emph{a})
in the above figure corresponds to Fig.\ref{fig:first}(\emph{c})
and the right panel (\emph{b}) above corresponds to Fig.\ref{fig:first}(\emph{d}).
These results are plotted at the {end} of $\tau=100\equiv\varepsilon^{3/2}t$.}\label{fig:fct-1}
\end{figure}

\subsubsection{Comparison with {pulse-induced nonlinearities}}

Let us now consider a very recent experimental observation of large-amplitude
perturbation in a PINI plasma by Pathak and Bailung \citep{pathak2025}
(P25 thereafter), where a positive pulse is induced externally to
a PINI plasma resulting in a large-amplitude positive perturbation
(up to $70\%$ of equilibrium plasma density), which is then used
to study the effect of controlled Landau damping on nonlinear IAW.
This produces a dissipative shock front when dissipation (through
Landau damping) becomes large.
\begin{figure}
\begin{centering}
\includegraphics[width=0.5\textwidth]{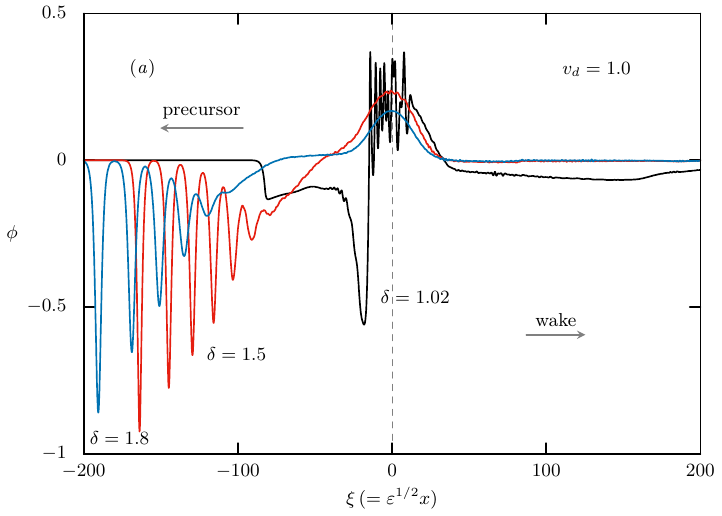}\hfill{}\includegraphics[width=0.5\textwidth]{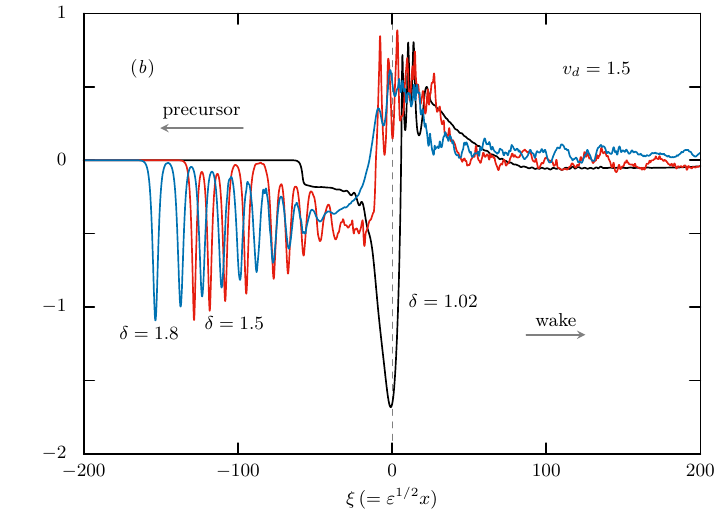}
\par\end{centering}
\caption{FCT simulation results of propagation of nonlinear wave in a PINI
plasma with {an} embedded external negatively charged debris. These
results are equivalent to the theoretical cases obtained through fKdV
solution shown in Fig.\ref{fig:second}. The left panel (\emph{a})
in the above figure corresponds to Fig.\ref{fig:second}(\emph{b})
and the right panel (\emph{b}) above corresponds to Fig.\ref{fig:second}(\emph{d}).
These results are plotted at the {end} of $\tau=100\equiv\varepsilon^{3/2}t$.}\label{fig:fct-neg}
\end{figure}

We argue that dynamically, an external perturbation can be considered
to be having the same effect as an external charged debris so far
as the leading edge (or precursor region) is concerned. So, we can
expect a similar behavior of nonlinear wave front for this external
perturbation and perturbation induced by an external debris. In this
context, we would like to compare our results as shown in the cases
for $\delta=1.02$ with $v_{d}=1.0,1.1$ in Fig.\ref{fig:first}(\emph{b})
and (\emph{c}) in the precursor regions and the case shown in Fig.4
of P25, which is comparable to our theoretical situation. In both
cases, we have a leading shock front with oscillations which are nothing
but DSWs when there is no dissipation. Note that Fig.4 of P25 is for
the case when Landau damping is negligible with negligible negative
ions $(\delta\sim1.0)$. The situation is however different in the
wake region (or trailing region of the pulse). This is due to the
fact that experimentally a one-time external perturbation is induced
by pushing a signal which then propagates in the form of discontinuity
(either as a single {soliton-like pulse or a ramp structure}) through the plasma
which then transforms into a DSW as it moves. For the case of external
debris however, the perturbation remains intact all throughout the
life cycle of the nonlinear wave front giving rise to a very definitive
signature in the wake region.

\subsection{{Negatively} charged external debris}

We now present the results for a negatively charged debris in a flowing
PINI plasma in Fig.\ref{fig:second}. All the parameters are kept
same as in the case for positively charged debris. The shaded regions
in all the panels show the extent of the charged debris with the centered
dashed line indicating the debris position. As expected, bright pinned
solitons are formed for $\delta$ close to unity at $\delta=1.02$,
when the relative velocity between debris and PINI plasma $v_{d}$
exceeds a certain value {as shown in} Fig.\ref{fig:second}(\emph{a})
and (\emph{b}). It can be clearly seen that as $v_{d}$ increases,
these pinned solitons finally {vanish} even for $\delta\rightarrow1$,
{which is consistent with our observations about pinned
solitons in an $e$-$i$ plasma }\citep{kumar2016,das2024}. For higher
values of $\delta,$ IAWs in the precursor region propagate showing
an oscillatory shock excitation which is not fully formed at a lower
debris velocity. However as $v_{d}$ increases, amplitude of the precursor
wave-front also increases and fully developed DSWs are observed to
be formed. Here, it is worth mentioning that, similar behavior was
also reported for the propagation of a rarefactive pulse in presence
of positive ion beam in high negative ion density plasma \citep{bailung2010}{, where increasing the beam velocity resulted in a rise in soliton amplitude and a knee formation at the downstream, culminating in the formation of a shock-like structure.} A realistic fluid simulation for much longer
time period shows that these DSWs, as they propagate away from the
site of the debris, get transformed into a train of {solitary waves} after segregation of the peaks, about which we shall discuss
in Section IV. The pinned solitons at the site of the debris are deformed
due to the presence of the shock. Also, behind the debris, soliton-like
structures are seen to be excited due to ion-acoustic oscillations.
\begin{figure}
\centering{}\includegraphics[width=0.5\textwidth]{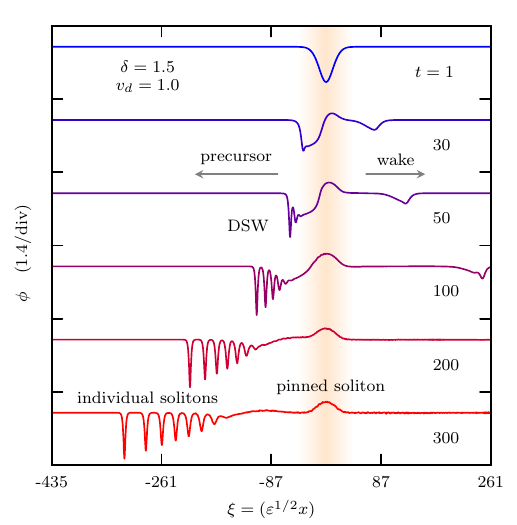}
\caption{FCT simulation results showing evolution of a DSW in the precursor
region over time for a negatively charged debris perturbation. Shaded
region corresponds to the perturbation site.}\label{fig:fct-trans}
\end{figure}

\section{Flux-corrected transport (FCT) simulation}

To corroborate the theoretical analysis {presented
in Section III}, we now present an FCT simulation of the PINI plasma
with an embedded external charged debris.\foreignlanguage{english}{
We use a 1-D multi-fluid FCT code (\emph{m}FCT) \citep{sarkar2023}
based on Boris's original algorithm \citep{boris1973} with Zalesak's
flux limiter \citep{zalesak}. The FCT formalism requires the hydrodynamic
equations Eq.(\ref{eq:cont}-\ref{eq:n-}) to be put in the form of
a generalized continuity equation,
\begin{equation}
\frac{\partial f}{\partial t}=-\frac{\partial}{\partial x}(fv)+\frac{\partial s}{\partial x},\label{eq:fct}
\end{equation}
where $f=(n_{+},n_{-},n_{+}v_{+},n_{-}v_{-})$ is the physical quantity
to be solved, $(fv)$ is the corresponding flux, and $s$ is the source
term. The external charged debris is entered through the Poisson equation
Eq.(\ref{eq:pois}) as before which is to be solved at every time
step along with Eq.(\ref{eq:fct}).}{{} We must however
emphasize that the FCT simulation results contain both KdV-type and
NLSE-type nonlinear solutions with the latter being less dominant.
Nevertheless, in certain parameter regimes, we might see a considerable
deviation of the simulation results from fKdV solutions.}

Our FCT simulation results are presented through Fig.\ref{fig:fct-1}
for the case with positively charged debris and through Fig.\ref{fig:fct-neg}
for negatively charged debris. In Fig.\ref{fig:fct-1}, we have shown
the results of the FCT simulation for the fKdV cases shown in Fig.\ref{fig:first}(\emph{c}),
corresponding to Fig.\ref{fig:fct-1}(\emph{a}) and \foreignlanguage{american}{Fig.\ref{fig:first}(\emph{d})},
corresponding to Fig.\ref{fig:fct-1}(\emph{b}). These curves are
plotted at the end of $\tau=100\equiv\varepsilon^{3/2}t$. The $x$-axis
is also respectively scaled for $\xi\equiv\varepsilon^{1/2}x$. Our
estimate shows that the effective $\varepsilon\sim0.7$ for $\rho_{0}=0.5$,
which is used in the fKdV solution as well as in the FCT simulation.
In both cases, the external charged debris is a Gaussian profile given
by $\rho_{{\rm ext}}(\xi)=\rho_{0}e^{-\xi^{2}/\Delta}$.
As we can see that the { simulation results} are quite
comparable to those obtained from the corresponding fKdV solutions,
except in the case for $\delta=1.5$ and $v_{d}=1.5$, which can be
attributed to slightly different {times} over which various nonlinearities
develop in a simulation and in fKdV solutions.
\begin{figure}[H]
\begin{centering}
~\includegraphics[width=0.495\textwidth]{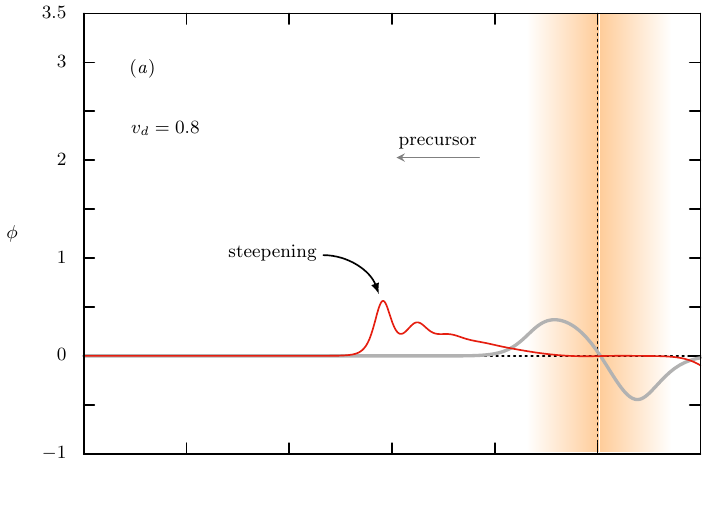}\hfill{}\includegraphics[width=0.495\textwidth]{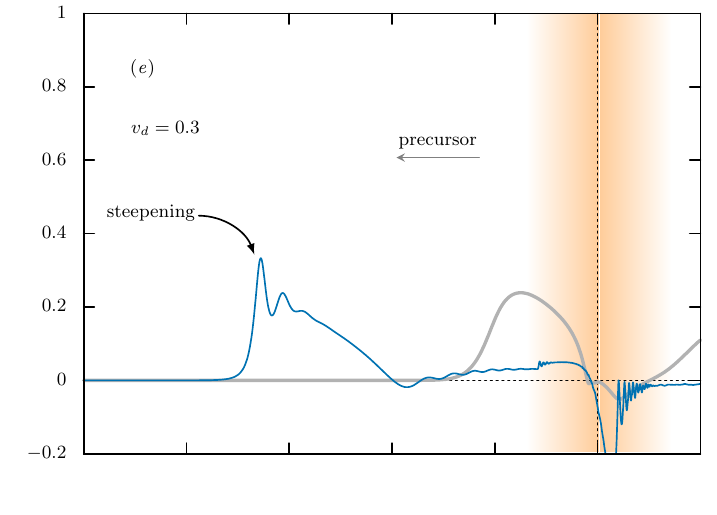}\\
~\vskip-32pt~\includegraphics[width=0.495\textwidth]{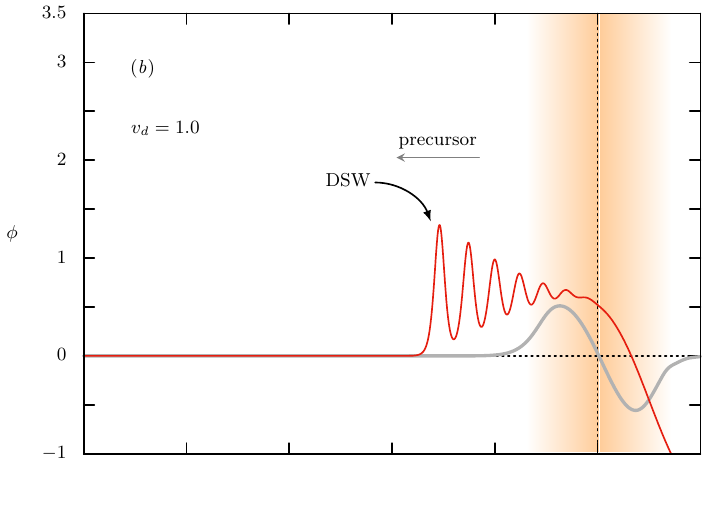}\hfill{}\includegraphics[width=0.495\textwidth]{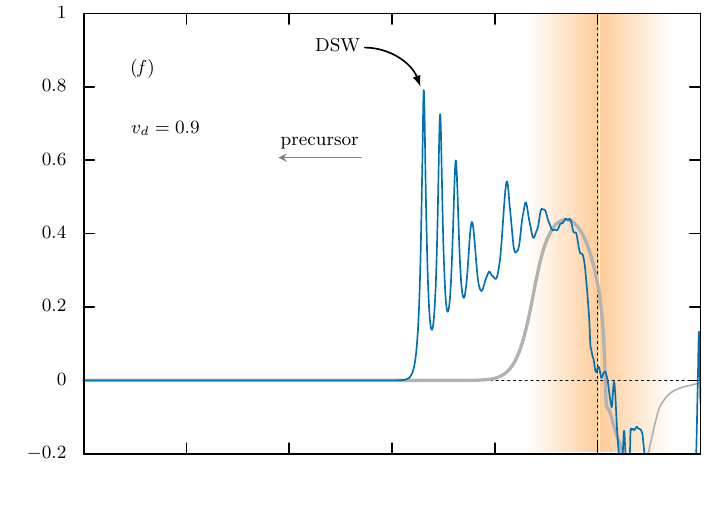}\\
~\vskip-32pt~\includegraphics[width=0.495\textwidth]{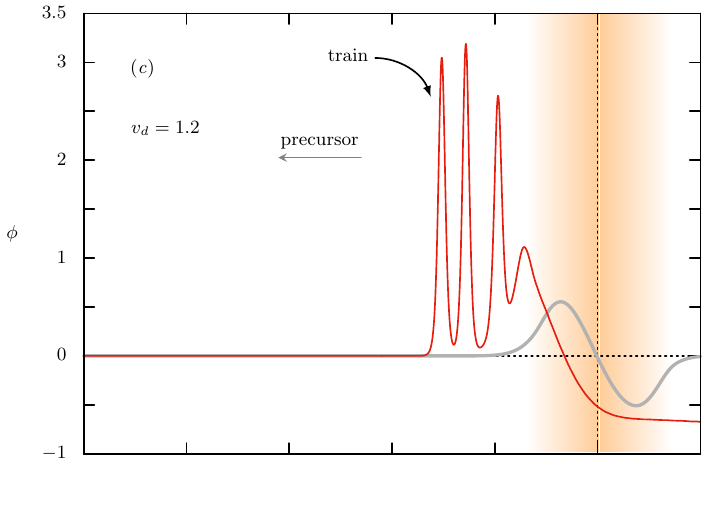}\hfill{}\includegraphics[width=0.495\textwidth]{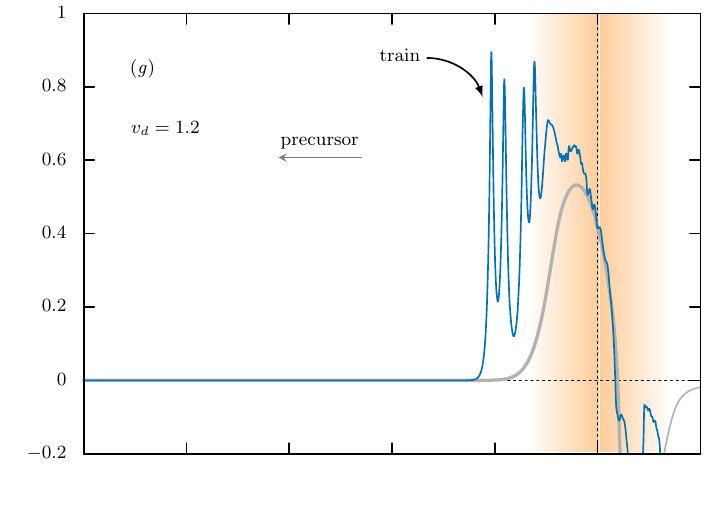}\\
~\vskip-32pt~\includegraphics[width=0.495\textwidth]{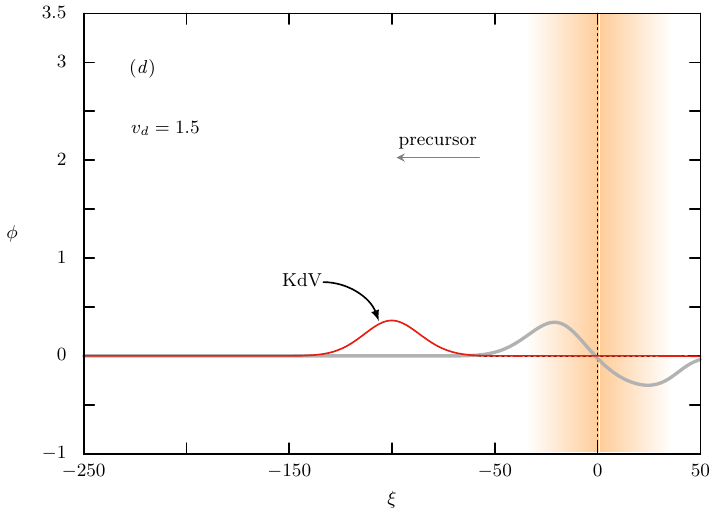}\hfill{}\includegraphics[width=0.495\textwidth]{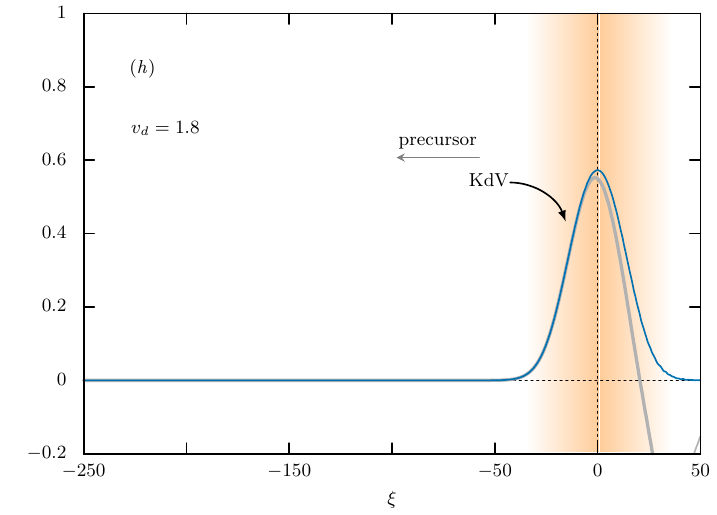}
\par\end{centering}
\caption{The changing nonlinearity with increasing debris velocity (from top
to bottom) as obtained from the fKdV solutions (panels (\emph{a\/})-(\emph{d\/}) correspond to $v_d=0.8,1.0,1.2$, and $1.5$, respectively) and FCT simulation (panels (\emph{e\/})-(\emph{h\/}) correspond to $v_d=0.3,0.9,1.2$, and $1.8$, respectively). The thick, gray pulse in each frame is the debris-induced perturbation at the initial stage of the evolution
which has later transformed into their respective form as time progresses.}\label{fig:nonlin}
\end{figure}

\selectlanguage{american}%
We now present the simulation {results} for negatively
charged debris in Fig.\ref{fig:fct-neg}. All other parameters except
the debris charge is same as in the case of Fig.\ref{fig:fct-1}.
Fig.\ref{fig:fct-neg}(\emph{a}) corresponds to the fKdV solution
presented in Fig.\ref{fig:second}(\emph{b}) $(v_{d}=1.0)$ and Fig.\ref{fig:fct-neg}(\emph{b})
corresponds to Fig.\ref{fig:second}(\emph{d}) $(v_{d}=1.5)$. {We
observe that} bright pinned solitons for $\delta=1.02$ {and
DSWs for higher $\delta$ are} properly captured in the FCT simulation,
while the DSWs seem to be fully formed {in the fKdV
solutions} only when $v_{d}=1.5$. Besides, the distinct soliton-like
structures in the wake regions of the fKdV solutions for both $v_{d}=1.0$
and $1.5$ seem to be absent in the simulation. Instead, singly peaked
pinned solitons with almost no wakes for $v_{d}=1.0$ and peak-modulated
pinned solitons \citep{kumar2016} and a turbulent wake trail due
to the propagating shock for $v_{d}=1.5$ are found to exist.
These
discrepancies between the fKdV solution and FCT simulations perhaps
can be attributed to the weak nonlinearity approximation that is in-built
in the fKdV derivation, while in FCT simulation there is no approximation
on nonlinearities involved. As a result, nonlinear oscillations and
structures tend to be more prominent in the simulation,{{}
which might also contain modulated wave-packet-like solutions (forced
NLSE type)}. Nevertheless, we believe that to large extent the fKdV
dynamics is similar to the FCT simulations and our observations about
similarities between debris-induced nonlinearities and nonlinearities
introduced by positive ion-beam and large-amplitude perturbation in
a PINI plasma can be fairly ascertained.

It is important to note that the DSWs seen in Fig.\ref{fig:second}
and \ref{fig:fct-neg} are called `quasi-steady state shocks' \citep{nakamura2004}
as the initial oscillatory shock profile is not retained over a longer
time period ($t\geq200$). Fig.\ref{fig:fct-trans} shows the potential
distribution of one such shock profile at different time-steps $(t=0-100)$.
Initially, the external charge perturbation introduces a highly nonlinear
IAW and the leading edge of the precursor becomes steepened which
forms a DSW with an oscillatory tail behind due to its dispersive
nature. This non-steady shock profile, after a relatively long time
$(t\geq200)$, transforms into a train of individual propagating {solitary waves}
as nonlinearity gets balanced by dispersion and each and every oscillatory
peaks in the DSW separates into individual {solitary wave} structures. This
effect is already observed and confirmed by earlier investigations
on ion-acoustic shock waves both in theory and experiments \citep{nakamura2004}.
The positive hump observed at the site of the debris is the signature
formation of a single-peak bright pinned soliton.

\section{Debris velocity and nonlinearity}

One very interesting phenomenon is the changing property of the nonlinearity
with increasing debris velocity which is similar to what is observed
experimentally with respect to beam velocity in a PINI plasma \citep{sharma2010}.
{From our analysis}, what we observe is that as debris
velocity increases, the wave amplitude steadily increases until reaching
a maximum and then decreases, which is seen from the steepening in
the leading edge (precursor region), which then gradually changes
to a DSW before becoming a train of {solitary waves} and then a single {KdV
soliton-like structure}. While the steepening in the leading edge indicates a dominant
positive nonlinearity over dispersion at low debris velocity, as debris
velocity increases, nonlinearity decreases and gets balanced by dispersion
which creates a {KdV soliton-like structure.}
In Fig.\ref{fig:nonlin}, we show
this effect as obtained from the fKdV solution {[}left column Fig.\ref{fig:nonlin}(\emph{a}-\emph{d}){]}
and from FCT hydrodynamic simulation {[}right column Fig.\ref{fig:nonlin}(\emph{e}-\emph{h}){]}.
In this figure, we have indicated the changing nonlinearity with an
arrow with the keywords `steepening', `DSW', `train', and `KdV', respectively
to indicate the phenomenon of steepening in the leading edge, formation
of DSW, transformation to a train of {solitary waves}, and finally a simple
KdV {soliton-like} pulse as debris velocity $v_{d}$ increases from $0.8$ to $1.5$
for fKdV solutions and from $0.3$ to $1.8$ for FCT solutions.
All the panels{{} in the above figure }are drawn at
the rest frame of the debris, which is moving toward left. The thick,
gray colored pulse in each frame indicates the initial stage of the
perturbation induced by the moving debris which has later transformed.{{}
Density ratio $(\delta)$ and ion to electron temperature ratio $(\sigma)$
are taken to be $1.05$ and $0.05$ respectively, in both cases}.
It is important to note that all these evolutions are drawn at the
same time evolution (column wise) across different $v_{d}$. The parameters
are slightly different for fKdV and FCT solutions which are optimized
respectively to represent this effect. This parameter difference of
fKdV and FCT solutions can be attributed to the weak nonlinearity
approximation that is inherent in the KdV formalism and full hydrodynamic
simulation through FCT simulation.

We should also note that all these nonlinearities change their behaviors
as the plasma evolves in time. For example, a steepening of the leading
edge will eventually transform itself into another form as we evolve
the system in time. This is also evident from the similar experimental
observations \citep{sharma2010,bailung2010}. In this context we would
like to recall that a pure KdV equation is integrable with `${\rm sech}$'
solution which \emph{does not} get evolved in time. However, an fKdV
equation is inherently non-integrable and different nonlinear solutions
of the fKdV equation such as DSW or pinned solitons are necessarily
transient phenomena and are bound to evolve in time. As per the parameters
used in the FCT simulation, assuming an equilibrium plasma number
density of $\sim10^{6}\,{\rm cm}^{-3}$, the entire spatial nonlinear
evolution frame shown in Fig.\ref{fig:nonlin} is contained within
$\sim25\,{\rm {\rm cm}}$, which is quite comparable to distance over
which these phenomena are observed in contemporary plasma devices
\citep{sharma2010,bailung2010}.

\section{Summary and conclusion}

To summarize, we have presented a detailed study of external debris-induced
nonlinear structures in a multicomponent positive ion-negative ion
(PINI) plasma. The motivation to choose a PINI plasma for this study
is primarily {due to the} fact that PINI plasmas are easily produced in
low-temperature laboratory devices and also are widely used in plasma
processing experiments. {Additionally}, theoretical
and experimental (laboratory and also space-based) studies of debris-induced
nonlinear plasma phenomena have been recently generating a lot of
interests among the plasma physicists, especially due to the renewed
attention of the scientific community in detection and removal of
space-debris from low-earth orbits. Besides a detailed analysis of
nonlinear waves excited by external charged debris, we also point
out the similarities in certain characteristics with those excited
by a high velocity positive ion beam and large amplitude pulse perturbations
in such plasmas. 

{The nonlinear wave structures observed in this study are a combined result of several physical mechanisms, governed by the response to the external debris perturbation by the plasma. Nonlinear precursor and wake waves are generated and propagated as an effect of the plasma ions in the vicinity of the perturbation going through ion-acoustic oscillations. The nature of debris charge also has a prominent effect on the plasma response\cite{sarkar2023,das2024}. For instance, a positively charged debris leads to the formation of an ion hole at the location of perturbation and at higher debris velocities, the excited precursor waves transform into oscillatory DSWs. Whereas, negative debris introduces IICSI forming phase-space vortices at the debris site, where some ions get trapped which leads to the formation of pinned solitons. However for a PINI plasma, the character of the excited nonlinear structures, whether compressive (bright) or rarefactive (dark), is intrinsically determined by the concentration of the negative ions present. When there is a very small amount of negative ions present, the dynamics gets completely dominated by positive ions, and in the presence of a positive debris, usual ion-hole formation occurs that leads to the bright DSW formation in the precursor region. However, in the presence of higher amount of negative ions, IICSI-like phenomena occurs for negative negative ions, where, negative ions get attracted to the positive debris leading to the formation of dark pinned solitons. Opposite phenomena occurs when the debris is considered to be negatively charged and structures also get modified accordingly.} The important findings of our analysis are presented
below.

\subsubsection*{Debris-induced nonlinear waves}
\begin{itemize}
\item For positively charged debris moving with subsonic velocities, small
amplitude bright {solitary waves} are observed with steepening in the leading
edge in low $\delta$ (low negative ion density) plasmas. On the other
hand, steepening starts on the falling edge of the wave as $\delta$
is increased beyond {a certain} value and as debris velocity reaches {the} sonic
and supersonic regime, large amplitude dark pinned solitons are excited.
However, for the same velocity regime, DSWs are seen to be formed
in low $\delta$ plasma. At very high debris velocity, the precursor
wave is found to transform into a KdV  {soliton-like structure.}
\item On the contrary, for perturbation induced by a negatively charged
debris, the nature and morphology of nonlinear structures are exactly
opposite to what is observed in the case of positively charged debris.
Presence of higher concentration of negative ion species (high $\delta$)
leads to formation of quasi-static DSWs as debris velocity becomes
supersonic and bright pinned solitons are formed in the presence of
lower concentration of negative ions (low $\delta$). However, the
nature of nonlinearity of the excited waves remains the same as in
the case of a positive debris perturbation.
\item Toward the end, we have also shown how the KdV nonlinearity gets modified
by the nonlinearity introduced by the external debris as debris velocity
increases. This observation also has {a similarity} with the phenomenon
of changing property of the nonlinear wave resulting out of beam-plasma
interaction as beam velocity increases \citep{nakamura1999,sharma2010,bailung2010}.
\end{itemize}

\subsubsection*{Can other {processes} mimic debris-induced nonlinearity?}

We have argued that a high velocity ion beam would behave analogous
to {a moving} external charge debris as long as the ion-acoustic timescale
is smaller than the electron-ion collision time $(\tau_{{\rm IA}}\ll\tau_{ei})$
or beam ions move faster without remaining in phase with plasma waves,
thus hindering beam plasma energy transfer process and would affect
nonlinear wave dynamics via the electric field, similar to an external
charge perturbation.
\begin{itemize}
\item Toward this, we observe a resemblance between the DSWs excited by
moving debris in a low negative ion density PINI plasma to the oscillatory
IA shocks excited by positive ramp signal in absence of the negative
ions \citep{pathak2025}.
\item Similarly, DSWs are observed to be excited by negatively charged debris
in a high $\delta$ plasma when debris velocity crosses the sound
speed which are also observed in negative pulse-excited high negative
ion density plasma with a positive ion beam \citep{bailung2010}.
\item We have also observed that positively charged debris introduce a positive
nonlinearity when the negative ion concentration is low (low $\delta$),
while the same debris excites a negative nonlinearity when the negative
ion concentration is high (high $\delta$). This phenomenon is similar
to what is observed with a positive ion beam in a PINI plasma \citep{nakamura1999}.
\item {Further,} the amplitudes of nonlinear waves are observed to increase
for both positive and negative perturbation with increase in the debris
velocity, which is similar to the case of increasing beam velocity
in PINI plasma \citep{nakamura1999,sharma2010,bailung2010}.
\end{itemize}
{It this context, we would like to draw a parallel between the nonlinear steepening that we have observed as the debris velocity increases and behavior of `density-divergence' found to exist in the propagation of solitary wave in a cold collision-free plasma along the magnetic field \citep{mach}. This threshold-driven transition - debris velocity in our work (See Figs.\ref{fig:first} and \ref{fig:fct-trans}) and Mach number in the latter \citep{mach} is fundamental to these nonlinear  waves despite their widely different plasma conditions. While our work is a purely electrostatic one and driven by external charged debris, the other one \citep{mach} investigate solitary wave in a cold plasma along a magnetic field. 
}

{This work advances the understanding of externally
driven nonlinear wave dynamics in multicomponent plasmas, offering
a broader perspective on the applicability and limitations of perturbative
approaches. The observed similarities between debris-driven nonlinear
structures {and} those due to other processes provide insights into
the underlying mechanisms and may help in the interpretation of future
experimental results involving negative ion plasmas.}
\begin{acknowledgments}
{The authors would like to thank Kalpana Bora for
critical reading of the revised manuscript and the anonymous referees
for their suggestions.} One of the authors, HS thanks CSIR-HRDG, New
Delhi, India for Senior Research Fellowship (SRF) research grant 09/059(0074)/2021-EMR-I.
\end{acknowledgments}

{
\section*{Appendix A: Weak nonlinearity and fKdV formalism}

{\global\long\def\theequation{A\arabic{equation}}%
\setcounter{equation}{0} 

Going by the expansions used in the derivation of the fKdV equation, Eqs.(\ref{eq:nexp}-\ref{eq:phiexp}), the expansion parameter at any instant of time can be written as
\begin{equation}
\varepsilon\sim\max\left(\left|\phi^{(1)}\right|,\left|n_{\pm}^{(1)}\right|,\left|v_{\pm}^{(1)}\right|\right).
\end{equation}
However, in our case, this condition seems to be satisfied only in the initial phase and as the solution evolves in time, it definitely goes out of the weak nonlinearity regime and enters a fully nonlinear territory, where the reductive perturbation theory is \emph{not} valid strictly. And yet, the fully nonlinear numerical solution of the system as obtained from the \emph{m}FCT code closely resembles that of the fKdV solution.

This can be understood from the global property of the KdV-type solutions where nonlinearity and dispersion tend to balance each other resulting in a soliton. In this case, despite the external force term due to the charged debris, the global solution of the fKdV equation still maintains an equilibration between the leading-order nonlinearity (the quadratic $\phi$ term)  and the leading-order dispersion (the third derivative term) preserving the overall nonlinear structure, resembling the simulation results. Toward this, we would like to refer to Infeld and Rowland \citep{infeld}, who have shown that KdV-like solitons persist and accurately describe the full nonlinear system even when the perturbation amplitude is not very small and KdV-like models can work way beyond their formal weak nonlinearity (small $\varepsilon$) limit. Similarly, Boyd \citep{boyd} explains that KdV-like equations often remain remarkably accurate even when $\varepsilon\sim{\cal O}(1)$, especially for localized initial conditions (viz.\ an external debris) and smooth evolution (smaller time step).
}}


{
\section*{Appendix B: External charged debris as first order perturbation}

{\global\long\def\theequation{B\arabic{equation}}%
\setcounter{equation}{0} 

The reductive perturbation theory employed to derive the classical KdV equation essentially stems out of the linear theory. {For} a simplified $e$-$i$ plasma with no external debris, the linear dispersion relation can be written as
\begin{equation}
\omega^2=k^2\left(\frac{1}{1+k^2}+\sigma\right),\label{eq:lin}
\end{equation}
where $\omega$ and $k$ are the frequency and wave number  of perturbation with the linear perturbation defined as $\sim e^{-i\omega t+ikx}$. Note that in the above equation, the same normalization is used as in Section II. With an external debris having charged density $\rho_{\rm ext}$, the dispersion relation becomes
\begin{equation}
k^2\left(1+\frac{1}{k^2\sigma-\omega^2}\right)+1=\frac{\rho_{\rm ext}}{\phi_1},
\end{equation}
where $\phi_1$ denotes the first order perturbation in plasma potential. This reduces to Eq.(\ref{eq:lin}) for $\rho_{\rm ext}=0$. So, one can readily see that a physically relevant and correct linear dispersion relation exists only when $\rho_{\rm ext}\to0$ at the first order or $\rho_{\rm ext}$ itself is expressed in terms of first order potential $\phi_1$. However, physically as $\rho_{\rm ext}$ is a purely external terms with no relation to plasma parameters, the latter possibility is ruled out leaving the only option of setting $\rho_{\rm ext}=0$ at the first order.

As the reductive perturbation theory is a nonlinear extension of the linear theory, this is also the precise reason why the reductive perturbation theory necessarily requires all first order external perturbations to vanish and any such perturbations must exist only at ${\cal O}(2)$ or higher.}}


{
\section*{Appendix C: KdV equation with linear Landau damping}}

{\global\long\def\theequation{C\arabic{equation}}%
\setcounter{equation}{0} In this Appendix, we try to estimate the Landau damping rate for a
simplified electron-heavy ion plasma where we have neglected the negative
ion species. For this case, using a reductive perturbation theory
along with the kinetic correction due to linear Landau damping of
the IAW, we have a KdV-like equation \citep{vandam}
\begin{equation}
\partial_{\tau}\phi^{(1)}+\frac{\beta}{\alpha}\phi^{(1)}\partial_{\xi}\phi^{(1)}+\frac{1}{\alpha}\partial_{\xi}^{3}\phi^{(1)}+\tilde{\gamma}\cdot{\cal P}\int_{-\infty}^{+\infty}\frac{\partial\phi^{(1)}}{\partial\xi'}\frac{d\xi'}{\xi-\xi'}=0,\label{eq:van}
\end{equation}
where
\begin{eqnarray}
\lambda & = & 1+\frac{3}{2}\sigma,\\
\alpha & = & 2\lambda^{-2}-3\lambda^{-4}\sigma,\\
\tilde{\gamma} & = & \frac{\lambda}{\alpha\varepsilon\sqrt{2\pi}}\left[\left(\frac{m_{e}}{m_{+}}\right)^{1/2}+\sigma^{-3/2}\exp\left(-\frac{\lambda^{2}}{2\sigma}\right)\right],
\end{eqnarray}
and ${\cal P}$ denotes the principal value integration. The last
term with the principal value integration in Eq.(\ref{eq:van}) contains
the linear Landau damping term of  associated IAW and the Eq.(\ref{eq:van})
as a whole accounts for weakly nonlinear IAW with Landau damping.
We note that integration in Eq.(\ref{eq:van}) is in the form of a
convolution integral (with of course the principal value) and can
be written as
\begin{equation}
{\cal P}\int_{-\infty}^{+\infty}\left(\frac{\partial\phi}{\partial\xi'}\right)\frac{d\xi'}{\xi-\xi'}={\cal P}\left(\frac{1}{\xi}\right)*\left(\frac{\partial\phi}{\partial\xi'}\right),
\end{equation}
where ${\cal P}(\xi^{-1})$ is the principal value distribution of
$\xi^{-1}$ and the star `$*$' denotes the convolution operation.
Taking Fourier transform ${\cal F}$ of the above expression and applying
convolution theorem, we have
\begin{equation}
{\cal F}\left[{\cal P}\int_{-\infty}^{+\infty}\left(\frac{\partial\phi}{\partial\xi'}\right)\frac{d\xi'}{\xi-\xi'}\right]={\cal F}\left[{\cal P}\left(\frac{1}{\xi}\right)\right]{\cal F}\left[\left(\frac{\partial\phi}{\partial\xi'}\right)\right].
\end{equation}
The Fourier transform of the terms on the right hand side can be readily
evaluated,
\begin{eqnarray}
{\cal F}\left[{\cal P}\left(\frac{1}{\xi}\right)\right] & = & -i\pi\cdot{\rm sgn}(k),\\
{\cal F}\left[\left(\frac{\partial\phi}{\partial\xi'}\right)\right] & = & ik\,\hat{\phi}(k),
\end{eqnarray}
where $\hat{\phi}(k)$ is Fourier transform of the function $\phi(\xi)$.}

{Considering now only the contribution of the Landau
damping term, neglecting the nonlinear and the dispersion terms of
Eq.(\ref{eq:van}), and taking Fourier transform in space and time
domain, we obtain the linear dispersion relation with only the Landau
damping term as
\begin{equation}
\omega=-i\cdot\tilde{\gamma}\cdot\pi k\cdot{\rm sgn}(k)\equiv-i\pi\tilde{\gamma}|k|,
\end{equation}
with $(\pi\tilde{\gamma}|k|)$ as the damping rate. Note that the
expression for damping rate can also be derived intuitively by comparing
the dimensions with that of the dissipation term in a KdV-Burgers
equation in a viscous plasma with viscosity coefficient $\nu$
\begin{equation}
\left[\tilde{\gamma}\cdot{\cal P}\int_{-\infty}^{+\infty}\frac{\partial\phi^{(1)}}{\partial\xi'}\frac{d\xi'}{\xi-\xi'}\right]=\left[\nu\frac{\partial^{2}\phi^{(1)}}{\partial\xi^{2}}\right],
\end{equation}
where the term on the right hand side is the viscous damping term
in a KdV-Burgers equation. Note that Eq.(\ref{eq:van}) is same as
the KdV-Burgers equation with last term replaced by the dissipative
damping term. The viscous damping rate of a fluid modeled by the KdV-Burgers
equation is given by $(\nu k^{2})$, with $k$ being the wave number
of the associated wave, which is in this case is IAW. So, by dimensional
analysis, we have}
\begin{equation}
[\nu k^{2}]\simeq[\tilde{\gamma}k].
\end{equation}



\end{document}